\documentclass[final,5p,times,twocolumn]{elsarticle}
\pdfoutput=1

\usepackage{graphicx} 
\usepackage{subfig}
\usepackage{enumerate}
\usepackage{amssymb, amsmath}
\usepackage[colorlinks]{hyperref}

\def\apj{{\it Astrophys.~J.}}

\def\aj{{\it Astronom.~J.}}
\def\apjl{{\it Astrophys.~J.~Lett.}}
\def\prd{{\it Phys.~Rev.~D}}

\def\mnras{{\it Mon.~Not. Roy.~Astr.~Soc.}}
\def\ijmpd{{\it Int.~J.~ Mod. Phys. D}}

\def\AnA{{\it Astron. Astrophys.}}
\def\grg{{\it Gen. Rel. Grav.}}
\def\ARAnA{{\it Ann. Rev. Astron. Astrophys.}}

\begin{document}

\begin{frontmatter}

\title{Testing the consistency between cosmological measurements of distance and age}

\author[CTP]{Remya Nair}
\ead{remya$_{-}$phy@yahoo.com}

\author[CTP]{Sanjay Jhingan\corref{cor1}}
\ead{sanjay.jhingan@gmail.com}

\author[DJ]{Deepak Jain}
\ead{djain@ddu.du.ac.in}

\address[CTP]{Centre for Theoretical Physics, Jamia Millia Islamia,
New Delhi 110025, India}

\address[DJ]{Deen Dayal Upadhyaya College,
University of Delhi, New Delhi 110015, India}

\cortext[cor1]{Corresponding author}

\begin{abstract}
We present a model independent method to test the consistency between 
cosmological measurements of distance and age, assuming the distance duality relation. We use type Ia supernovae, baryon acoustic oscillations, and
observational Hubble data, to reconstruct the luminosity distance $D_L(z)$, the angle averaged distance $D_V(z)$ and the Hubble rate $H(z)$, using Gaussian processes regression technique. We obtain estimate of the distance duality
relation in the redshift range $0.1<z<0.73$ and we find no evidence
for inconsistency between the data sets used.
\end{abstract}

\end{frontmatter}


\section{Introduction}

An explanation for the observed accelerated expansion of  the Universe still eludes cosmologists. The evidence for the existence of dark energy (DE), which is believed to source this acceleration \cite{de,sah,sami}, has been continuously piling since the first indications from supernovae (SNeIa) observations \cite{sne,accn}. Thanks to the abundance of good quality data cosmology has become a precision science. But even in this era of data driven cosmology we know little about constituents of our Universe. There are several large cosmological surveys that are aiming to answer some of the key question in cosmology today, like: what is the source of the acceleration of the Universe and what are the properties of this mysterious source. Further, there are many ambitious projects that are planned for the future \cite{surv}. In this scenario it is important to check whether the measurements coming from various surveys are consistent with each other. This is crucial since multiple probes are often combined to put better constraints on cosmological parameters. Considered together, one data set resolves the difficulties of the other, allowing certain degenerate parameters to be determined with far greater precision. A consistency check would also help us to examine different data sets for the presence of systematics.

Knox et al., examined consistency of the different cosmic
microwave background data sets to check whether the data are
contaminated by some residual non-cosmological signals \cite{kn}. Avgoustidis et al., studied the consistency among SNeIa, measurements of the Hubble parameter, and the baryon acoustic oscillation scale. Regarding the tension between the baryon acoustic oscillation scale and SNeIa data in light of possible deviations from transparency, they concluded that the source of the discrepancy may most likely be found among systematic effects of the modelling of the low redshift data, or it may be a statistical fluke \cite{avg}. Shafieloo et al., compared two different probes of the expansion history of the universe, namely, luminosity distances from SNeIa and angular diameter distance from galaxy clusters. They proposed a model independent method to search for inconsistencies between SNeIa and galaxy cluster data sets \cite{shaf}. More recently, Hazra and Shafieloo studied the consistency of the angular power spectrum data from WMAP and Planck
looking for possible systematics \cite{haz}. Cao and Zhu used
observational data with four angular diameter distance measurements and synthetic SNeIa+GRB observations for luminosity distance data, to investigate the tension between these two cosmological distances considering three classes of dark energy equation of state reconstruction. They found that the angular diameter distance measurements and the luminosity distance data are compatible at $1\sigma$ level \cite{cao}. Ruiz and Huterer tested the consistency of the standard $\omega$CDM model in the framework of General Relativity by separating information between the geometry and growth
of structure. Using data from SNeIa, baryon acoustic oscillations, the peak locations in the cosmic microwave background angular power spectrum, redshift space distortions, weak gravitational lensing and the abundance of galaxy clusters, and found that both geometry and growth separately favour the $\Lambda$CDM cosmology \cite{hut}.

In this work we look for consistency in distance and age measurement data sets using the distance duality (DD) relation. We use baryon acoustic oscillation
(BAO) and SNeIa data for distance measurements and observational
Hubble data as age measurement. The plan of the paper is as follows:  we begin with a brief overview of  DD relation in section \ref{DDR}, in section \ref{met} we discuss data sets used and a quick over-view of Gaussian Processes which is the methodology used here, we discuss our results in section \ref{res}.

\section{Distance duality relation}\label{DDR}
In 1933 Etherington proved the reciprocity relation in area
distances, between a source and an observer in relative motion with
each other \cite{eth}. This relation is valid for any curved
spacetime, and even basic symmetry assumptions like homogeneity and isotropy are not required. The relation holds as long as gravity is described by a metric theory, photons travel on null geodesics and the geodesic deviation equation is valid \cite{bk,uz}. If photon number is conserved this further reduces to a relation between the angular diameter distance and the luminosity distance \cite{dd}:
\begin{equation}
D_{L}=D_{A} (1+z)^{2}.
\end{equation}
This is termed as the DD relation and plays an
important role in galaxy cluster observations and lensing studies
\cite{ellisgrg}. Since DD is crucial to cosmological studies and
plays a key role in how galaxy observations are analysed, it is
important to check its validity. Since both the distances in the
duality relation are observable quantities it is possible to test
this relation. Now, there are many ways in which cosmic distances
are measured. One can look for sources that can be used as standard
candles for deriving luminosity distance, and standard rulers can be
used to derive angular diameter distance. SNeIa can be used as
standard candles since they have a peak luminosity that is tightly correlated with the shape of their light curves and hence they can
be calibrated. On the other hand, combined measurements of the
Sunyaev-Zeldovich effect and X-Ray analysis (SZE/X-ray) provides a
measure of the angular diameter distance to a cluster.
The Baryon acoustic feature in the matter clustering is another
independent distance indicator and can be used as a standard ruler.
As observed in previous works (\citep{hol,us1}), the constraints
on the DD obtained from galaxy cluster measurements, depend on the
assumptions of cluster geometry (spherical or elliptical). Hence
assuming DD to be true, one can use it as a probe of cluster
geometry. DD has also been used to constrain the cosmic opacity
between different redshifts.

One of the assumptions in DD, is the conservation of photon number. Hence, the temperature redshift relation, relating the observed and emitted temperature of the cosmic microwave background photons, which derives from DD, also
assumes photon conservation. The relation will be modified if this assumption was violated. There are many mechanisms that have been proposed in literature which give rise to such a violation, for example decaying vacuum cosmology, photon axion coupling etc.. In this regard, there have been several attempts to measure the cosmic microwave background temperature at different redshifts, using, for example, quasar absorption line spectra and this can
be used to test the validity of the temperature shift relation.
But since the uncertainties are large, more data is required to
put robust constraints. Testing the DD offers another way to
confirm photon conservation. Assuming there are some unclustered
sources of photon attenuation in the Universe, one can use DD
to put constraints on the difference in opacity between two
redshifts. Refer to \cite{avg,more,us2} for details of such studies.

\section{Methodology and data sets used}\label{met}
\subsection{Methodology}
As mentioned earlier, we use SNeIa, BAO and observational Hubble
data to constrain DD (details of the data sets given in \ref{data}).
Since SNeIa are expected to form from standard
explosion of a white dwarf, they are assumed to have homogeneous
light curve and uniform luminosity. Although there is an
intrinsic scatter in the peak luminosities of SNeIa, an empirical
correlation exists between the shape of the SNeIa light curve and
the SNeIa luminosity. Hence these candles can be standardized, and
are used as standard candles for estimating luminosity distances.

Observational Hubble data is obtained from the measurement of the
relative ages of passively evolving galaxies. The Hubble rate
depends on the differential age of the Universe as:
\begin{equation}
H(z)=-\frac{1}{1+z}\frac{dz}{dt},
\end{equation}
and hence the determination of $dz/dt$ gives an estimate of $H(z)$.
For this one has to look for the variation of ages, $\Delta t$, with redshift $\Delta z$. Collection of galaxy samples of passively
evolving galaxies with high quality spectroscopy, are used and
differential ages for the samples are computed. These are then used
as estimates of $dz/dt$ which eventually gives an estimate of $H(z)$.

BAO refers to a length scale in the distribution of the photons and
baryons. This scale is imprinted in the matter distribution due to
the stalling of sound waves in the plasma of the early Universe, and
hence they can be treated as cosmological standard rulers. Enough
BAO data has not been accumulated to separately measure the
tangential and radial components of the signal, and hence the
two distances. But it is possible to constrain an angle-averaged
clustering measurement, obtained from the combination of two spatial
dimensions orthogonal to the line of sight and one dimension along
the direction of sight, as defined below \cite{eis}:
\begin{equation}
D_V^3 = \frac{D_A^2 c z (1+z)^2}{H(z)},
\label{dv}
\end{equation}
where, $D_A$ is the angular diameter distance, $z$ is the redshift
and $H$ is the Hubble rate. 

If the DD relation holds,
then we know that the luminosity distance and the angular diameter
distance are related as $D_L =D_A (1+z)^2$. Let us assume
\begin{equation}
\eta = \frac{D_L}{D_A (1+z)^2},
\end{equation}
where $\eta = 1$ if the DD relation holds. Using this
we can rewrite (\ref{dv}) as
\begin{equation}
\eta=\frac{D_L (cz)^{1/2}}{(1+z) H^{1/2} D_V^{3/2}}.
\label{eta}
\end{equation}
Now, if the distance measurements and the age measurements are
consistent with each other, and the DD holds then
$\eta$ in (\ref{eta}) should be equal to $1$. A deviation from $1$
may imply the breakdown of one or more of the assumptions mentioned
earlier or it may indicate the presence of some systematic in the
data sets used. Many authors have studied the DD
relation using various data sets (\cite{restdd}). Constraints
obtained from using cluster data for the angular diameter
distance estimate depends on the galaxy cluster model assumed.
Since we are using BAO data for the angular diameter distance
estimate our constraints does not contain such biases. Another
source of error in the analysis of DD is that it is not always
possible to obtain a luminosity distance estimate and an angular
diameter distance estimate at the same redshift, and some kind of
redshift-matching criteria is adopted. In this work we reconstruct
all the observable quantities ($H(z)$, $D_V(z)$ and $D_L(z)$)  in
the redshift range of interest using a non-parametric method
called Gaussian Processes and then estimate $\eta$ using these
quantities.

As mentioned earlier we take $D_V(z)$ measurements from BAO data,
$D_L(z)$ measurements from SNeIa data, and $H(z)$ measurements from
the observational Hubble data obtained from differential ages of
passively evolving galaxies, to check the above relation. The
details of the data sets are given in the next section.

\subsection{Data}\label{data}

\begin{itemize}
\item To estimate the luminosity distance $D_L(z)$ at various
redshifts we use the distance modulus measurements from SNeIa
Union2.1 sample, as given in \cite{suz}. This sample contains
$580$ supernovae and spans $0.015 < z < 1.414$. In the data set,
the distance modulus is given in terms of the redshift, and this is
used to estimate the luminosity distance. The relation between
the distance modulus $\mu$ and the luminosity distance $D_L$ is:
\begin{equation}
\mu_B(z) = m_B - M_B = 5 \log_{10} \left(\frac{D_L(z)}{1Mpc}\right)
+25,
\end{equation}
where $M_B$ is the absolute magnitude of the source and $m_B$ is
the apparent magnitude ($B$ is for $B$-band).

\item We use eight measurements from BAO data compiled from
different groups in the redshift range $0.106<z<0.73$.
We use data from SDSS ($z=0.2, 0.35$ and $0.15$), 6dFGS($z=0.106$),
WiggleZ($z=0.44, 0.6,0.73$) and BOSS($z=0.57$) for estimating the
distance \cite{perc, beut, boss1, kaz, ross}.

\item We also use the observational Hubble data as
compiled in \cite{rat}. Note here that we only use those data
points which are obtained from the analysis of differential ages
of galaxies.
\end{itemize}

\subsection{Gaussian Processes}\label{gp}
A Gaussian Process (GP) is a collection of random variables, any
finite number of which have a joint Gaussian distribution
\cite{rasm}. Just like a Gaussian distribution is a distribution of
a random variable (characterized by a mean and a covariance), a GP
is a distribution over functions. It is characterized by a mean
function and a covariance matrix. In a regression analysis the aim
is to infer the relation between independent and dependent variables,
given some set of observations. In parametric regression we assume
some functional relation between the output and the input
$f(x,{\boldsymbol \theta})$, where ${\boldsymbol \theta}$
represents the set of model parameters, and regression requires
finding the values of ${\boldsymbol \theta}$ which best describe
the data. Usually, the chi-squared merit function is minimized to
obtain the best fit parameters. Similarly in GP regression, the
function $f(x)$ is represented as
\begin{equation}
f(x) \sim GP(\mu (x),k(x,x')),
\end{equation}
which means that the value of $f(x)$ at any point $x$, is a
Gaussian random variable with mean $\mu (x)$ and covariance
$k(x,x')$:
\begin{eqnarray}
\mu (x) &=& E(f(x)),\\
k(x,x') &=& E((f(x)-\mu (x))(f(x')-\mu (x'))).
\end{eqnarray}
There are many choices for the covariance functions: squared
exponential, spline, polynomial etc. Here we chose the commonly
used squared exponential function for its simplicity. The squared exponential covariance function is expressed as:
\begin{equation}
k(x,x')=\sigma_f^2 \exp \left(-\frac{(x-x')^2}{2l^2}\right).
\end{equation}
This covariance functions is parameterized by the two
parameters, $\sigma_f$ and $l$ (known as hyperparameters), which
represent the length scales in the GP. $\sigma_f$ controls the
variation in $f(x)$ relative to the mean and $l$ corresponds
to the correlation length along which the successive $f(x)$ values
are correlated. As desired, the covariance is maximum for variables
whose inputs are very close which is expected for smooth functions.
The matrix elements of the covariance matrix for the GP:
$K({\bf X,X})$, are given by
\begin{equation}
[{\bf K(X,X})]_{i,j}=k(x_i,x_j).
\end{equation}
Similar to the function $f(x)$, the data ${\bf y}$ can also be
represented using GP:
\begin{equation}
y \sim GP(\mu (x),k(x,x')),
\end{equation}
Now, given a set of inputs ${\bf X}$ (also called training
vectors), outputs ${\bf y}$ (the data set, also called target),
and the covariance matrix ${\bf K(X,X})$, our aim is to make
inference about the function $f(x)$ at some other points
${\bf \hat{X}}$. The joint probability distribution for the
data ${\bf y}$ and the reconstructed function ${\bf \hat{f}}$ is
given by
\[
\begin{pmatrix}
{\bf y} \\ {\bf \hat{f}}
\end{pmatrix}  \sim  N \left( \begin{bmatrix} {\boldsymbol \mu} \\ \hat{{\boldsymbol \mu}} \end{bmatrix},
\begin{bmatrix} {\bf K(X,X)+C} & {\bf K(X,\hat{X}})\\
{\bf K(\hat{X},X}) & {\bf K(\hat{X},\hat{X}}) \end{bmatrix} \right),
\]
where ${\boldsymbol \mu}$ and $\hat{{\boldsymbol \mu}}$ are the
assumed means (initial guesses) and ${\bf C}$ is the covariance
matrix of the data, which is diagonal if the data points are
uncorrelated. After some matrix algebra one can rewrite the joint probability distribution as (please see \cite{rasm} or \cite{usgp}
for more details of the calculations)
\begin{multline*}
P({\bf y},{\bf \hat{f}})=\frac{1}{(2\pi)^{p/2} |
{\bf \Sigma}_{11}|^{1/2}} \exp [-\frac{1}{2}({\bf y}-
{\boldsymbol \mu})^T {\bf \Sigma}_{11}^{-1}({\bf y}-
{\boldsymbol \mu})]\\
\frac{1}{(2\pi)^{q/2} |{\bf A}|^{1/2}} \exp [-\frac{1}{2}
({\bf \hat{f}}-{\bf a})^T {\bf A}^{-1}({\bf \hat{f}}-{\bf a})],
\end{multline*}
where
\begin{equation}
{\bf a}={\hat{\boldsymbol \mu}}+{\bf K(X,\hat{X}})^T
({\bf K(X,X)+C})^{-1}({\bf y}-{\boldsymbol \mu}),
\label{mean}
\end{equation}
\begin{equation}
{\bf A}={\bf K(\hat{X},\hat{X}})-{\bf K(X,\hat{X}})^T
({\bf K(X,X)+C})^{-1} {\bf K(X,\hat{X}})
\label{var}
\end{equation}
and
\begin{equation}
{\bf \Sigma}_{11}={\bf K(X,X)+C}.
\end{equation}
Here $p$ and $q$ are the number of points in ${\bf X}$ and
${\bf \hat{X}}$ respectively. The marginal distribution of
${\bf y}$ is given by
\begin{equation}
P({\bf y})=\int P({\bf y},{\bf \hat{f}}) d{\bf \hat{f}} =
\frac{1}{(2\pi)^{p/2} |{\bf \Sigma}_{11}|^{1/2}} \exp
[-\frac{1}{2}({\bf y}-{\boldsymbol \mu})^T {\bf \Sigma}_{11}^{-1}
({\bf y}-{\boldsymbol \mu})],
\label{margp}
\end{equation}
and the conditional distribution $P({\bf \hat{f}}|{\bf y})$ is
\[
P({\bf \hat{f}}|{\bf y})=\frac{P({\bf y},{\bf \hat{f}})}{P({\bf y})}=\frac{1}{(2\pi)^{q/2} |{\bf A}|^{1/2}} \exp [-\frac{1}{2}({\bf \hat{f}}-{\bf a})^T {\bf A}^{-1}({\bf \hat{f}}-{\bf a})],
\]
which implies that the reconstructed function ${\bf \hat{f}}
({\bf \hat{X}})$ has a Gaussian normal distribution given by
\begin{equation}
{\bf \hat{f}} \sim GP({\bf a},{\bf A}).
\end{equation}
Here $\sigma_f$ and $l$ are unknown parameters of the GP and
training of a GP involves selecting appropriate values for
these parameters. This is usually done by maximising the marginal
log-likelihood probability $\ln P(y)$ (from (\ref{margp}))
\begin{multline}
\ln P({\bf y})= - \frac{1}{2} ({\bf y}-{\boldsymbol \mu})^T [{\bf K(X,X)+C}]^{-1} ({\bf y}-{\boldsymbol \mu}) \\
-\frac{1}{2}
\ln |{\bf K(X,X)+C}|-\frac{p}{2} \ln 2 \pi . \label{gpe}
\end{multline}
Further, to do the analysis one needs to specify an input mean
function. We chose the initial mean function for all quantities
to correspond to a flat $\Lambda$CDM model with $\Omega_m = 0.3$,
and $H_0=70$ Km/sec/Mpc. The means are adjusted during the analysis and
replaced by posterior means suggested by preliminary runs.
Our redshift range of interest is $0.1<z<0.73$ (governed by the
BAO data range). We divide this redshift range into intervals of
$\Delta z=0.001$. This gives us $631$ target points ($z$ values
at which the functions are reconstructed). We then reconstruct
$H(z)$, $D_L(z)$ and $D_V(z)$ at these points using GP regression.
This implies that for each $z$ target point we have a mean and
a variance of the reconstructed functions given by (\ref{mean}) and (\ref{var}) respectively. Note that maximising (\ref{gpe}) is
an approximation, and can be used if the posterior for
${\boldsymbol \theta}$, is fairly well peaked. Since this
is not always guaranteed, in our analysis, we sample the
hyperparameter space and the probability distributions
of the reconstructed function are weighted by the posterior
distributions of the hyperparameters (in effect we marginalize
over the hyperparameters). Samples from these weighted
distributions of $H(z)$, $D_L(z)$ and $D_V(z)$ are used to
eventually construct $\eta(z)$ as given in (\ref{eta}).
Note here that $\eta(z)$ itself is not a GP, i.e. it is not
constructed with GP methodology. It is derived using
\eqref{eta} and the errors on $\eta(z)$
are obtained using error propagation (including covariances at
different redshifts).
\begin{figure}[ht]
\centering
\includegraphics[width=0.45\textwidth]
{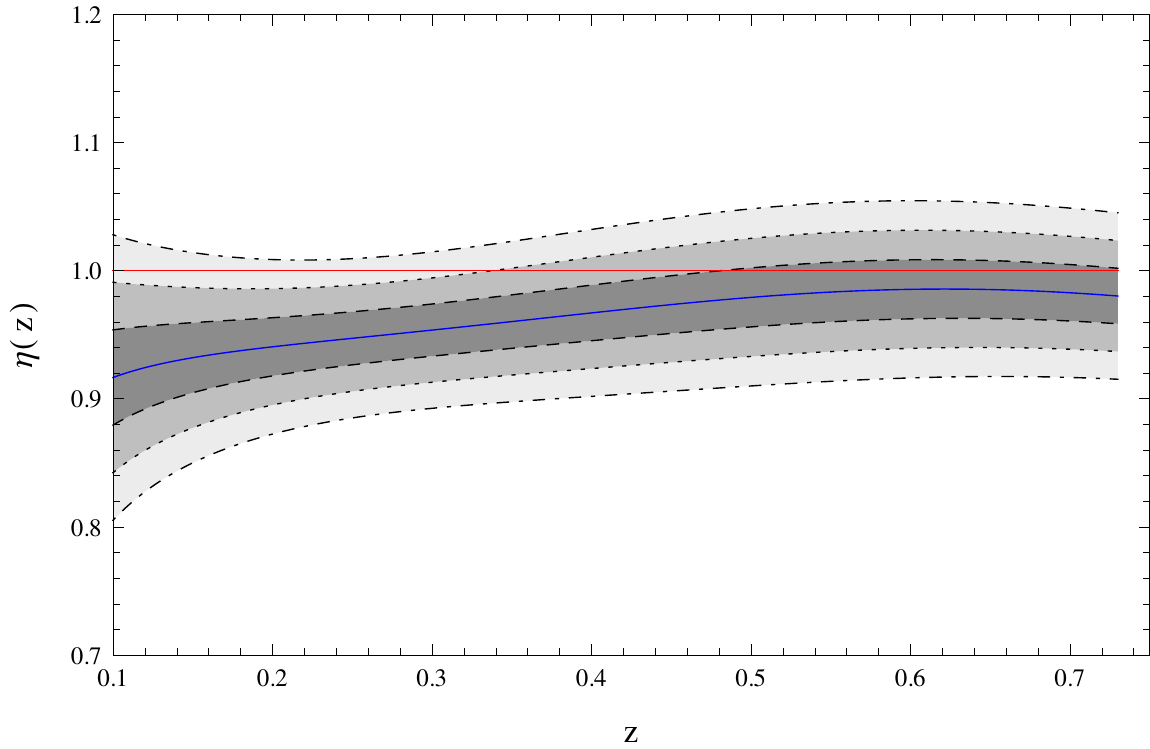}
\caption{Reconstructed $\eta$ (solid blue curve) as function
of $z$. The dashed, dotted and dot-dashed black curve represent $1$, $2$
and $3\sigma$ confidence levels respectively. Solid red line
represents $\eta=1$}
\label{etafig}
\end{figure}

\section{Results and Discussion}\label{res}
In this paper we have used DD relation as a check for consistency between different data sets. The main result of the paper is summarised in figure \ref{etafig}, where we plot the variation of
$\eta$ with redshift, as estimated from the GP reconstruction of
the luminosity distance, angle averaged distance and the Hubble
rate. The solid blue line is the best fit curve and the black curves
around it represent the confidence intervals. The red horizontal
line is the value of $\eta$ when DD holds (and all the
data sets are consistent with each other). 
Note that the error bars in this plot should be understood 
as point by point along the redshift. The estimates of the three
reconstructed quantities ($H(z)$, $D_L(z)$ and $D_V(z)$) have
correlations at different redshifts and so will the
estimates of $\eta(z)$, but these correlations are not visible
in this plot.

We observe that the
DD holds within 3$\sigma$ confidence level. These
model independent constraints are better than our previous
constraints on $\eta$ (\cite{us2}), where we took some simple parametrization for $\eta$ and fixed $H(z)$ assuming
$\Lambda$CDM cosmology. Further the mean value of $\eta$ is
slightly less than unity. This is similar to results obtained in
earlier works in this direction. Uzan et al. \citep{uz}, found a
best fit value for $\eta$ which was slightly less than one and
related this trend to the systematics in the SZE/X-ray analysis
of galaxy clusters, assuming $\Lambda$CDM cosmology. Bassett and Kunz \cite{bk}, in their three parameter model of DD violation, also found that the SNeIa sample they used were brighter relative to their $d_A$ data, . Gravitational lensing of the high-z supernovae was suggested as a possible explanation. Nesseris and Garcia-Bellido recently used Genetic Algorithm approach to extract model independent and bias-
free reconstruction information from SNeIa, BAO and the growth rate
of matter perturbations \cite{nes}. Our result is moderately
consistent with their analysis, but we do not recover the dip in
$\eta$ in the range $0.3<z<0.7$, that they obtain in their
reconstruction. $\eta$ can also be assumed to be a constant (other
than unity) and the value of the constant can be estimated from
observations, see for example one of our previous works \cite{us2},
Uzan et al., \cite{uz} or Bernadis et al \cite{bern}. 

Our model independent method can also be used to test DD relation
if the data sets are known to be consistent with each other. 
If this relation is found to be
inconsistent with observations, it would be a major problem for
observational cosmology, since the optical theorem that relates
surface brightness of an object at the source and observer, and
the temperature shift relation of the cosmic microwave background
are derived from this relation (\cite{ellisgrg}). In the event
that the DD relation is not valid, these key relations in cosmology 
would have to be modified. 
Future surveys (especially the increase
in BAO data points) would better constrain $\eta$ and this method
can be used to look for the presence of systematics within the data
sets.

\section*{Acknowledgments}
RN acknowledges support under CSIR-SRF scheme
(Govt. of India). DJ thanks A. Mukherjee and S. Mahajan for providing the facilities to carry out research, and CTP (JMI) for research support. SJ acknowledge support from grant under ISRO-RESPOND program (ISRO/RES/2/384/2014-15).

\end{document}